\documentclass[twocolumn,secnumarabic,amssymb, nobibnotes, aps, prd]{revtex4-2}
 
\usepackage[american]{babel}
\usepackage{graphicx}
\usepackage{dcolumn}
\usepackage{bm}
\usepackage{hyphenat}
\usepackage{amssymb}   
\usepackage{verbatim}
\usepackage{xpatch}
\usepackage{color}
\usepackage{epstopdf}
\usepackage{multirow}
\usepackage{mhchem} 

\usepackage{hyperref}

\begin{document}

\title{Ultrastable optical, XUV and soft-x-ray clock transitions in \\ open-shell highly charged ions}

\author{Chunhai Lyu}
\altaffiliation{chunhai.lyu@mpi-hd.mpg.de}
\affiliation{Max-Planck-Institut f\"{u}r Kernphysik, Saupfercheckweg 1, 69117 Heidelberg, Germany}
\author{Christoph H. Keitel}
\affiliation{Max-Planck-Institut f\"{u}r Kernphysik, Saupfercheckweg 1, 69117 Heidelberg, Germany}
\author{Zolt\'{a}n Harman}
\affiliation{Max-Planck-Institut f\"{u}r Kernphysik, Saupfercheckweg 1, 69117 Heidelberg, Germany}

\date{\today}

\begin{abstract}

Highly charged ions (HCIs) are insensitive to external perturbations and are attractive for the development of ultrastable clocks. However, only a few HCI candidates are known to provide optical clock transitions. In this Letter, we show that, as a result of strong relativistic effects, there are more than 100 suitable optical HCI clock candidates in more than 70 elements. Their transitions are embedded in the fine-structure splitting of the $nd^4$, $nd^5$ and $nd^6$ ground-state configurations with $n=3,4,5$ being the principal quantum numbers. The corresponding high multipolarity transitions in these ions have lifetimes and quality factors many orders of magnitude longer and larger, respectively, than those in state-of-the-art clocks. Their polarizabilities are also orders of magnitude smaller, rendering them more stable against external electromagnetic fields. 
Furthermore, within the same electronic configurations, the clock transitions in heavy ions scale up to the XUV and soft-x-ray region, thus enable the development of clocks based on shorter wavelengths. The existence of multiple clock transitions in different charge states of a single element, as well as in a whole isoelectronic sequence, would significantly enrich the detection of fine-structure constant variations, the search for new physics and the test of nuclear theories via high-precision spectroscopy.

\end{abstract}

\maketitle

%

High-precision optical clocks have wide applications in fields such as the new definition of time units~\cite{clockcomparison2021,clockcomparison2022a,clockcomparison2022b}, the test of variations of the fine-structure constant~\cite{alpha2004a,alpha2004b,alpha2014}, and the constrain of a conjectured fifth force~\cite{5force2018,Yb+5force2020,Ca+5force2020} and other new physics phenomena~\cite{newphysics-RMP2018,Lorentz2019}. 
The elements of these clocks are either neutral atoms or singly charged ions~\cite{ludlow2015optical} based on hyperfine-induced $ns^2~{}^1S_0-nsnp~{}^3P_0$~\cite{Al+-clock2019,Sr-lattice-clock2005,Cd-lattice-clock2019,In+-clock2019,Yb-lattice-clock2009,Hg-lattice-clock2012} and electric quadrupole $(n+1)s~{}^1S_{1/2}-nd~{}^2D_{5/2}$ transitions~\cite{Ca+-clock2016,Sr+-clock2004,Hg+-clock2001,CuAgAu-clock2021,Ra+-clock2022} with a quality factor $Q=\nu/\Delta\nu$ in the range of $10^{14}\sim10^{17}$. Here, $n$ is the principal quantum number, $\nu$ the transition frequency and $\Delta\nu$ the corresponding natural linewidth. For specific cases such as Lu$^+$ and Yb$^+$, electric quadrupole and octupole transitions between $6s^2-6s5d$~\cite{Lu+-clock2016,Lu+-clock2020} and between $4f^{13}6s^2-4f^{14}6s$~\cite{Yb+E3clock2012,Yb+E3clock2020} with $Q$ factor in the range of $10^{21}$ and $10^{23}$, respectively, are also employed.  

Nevertheless, these clocks suffer from large systematic shifts and uncertainties induced by black body radiation~\cite{BBR2018,Tm-clock2019} and external trapping fields~\cite{BBR2016,ludlow2015optical}. To overcome these difficulties, clocks based on a nuclear transition in the $^{227}$Th isomer~\cite{NuclearTh-clock2012,NuclearTh-clock2019,NuclearTh-clock2021} and electronic transitions in highly charged ions  (HCIs)~\cite{kozlov2018highly,HCIclock-Ar13-2022,EUVclock-lyu,HCIclock-2022} are put forward as they bear smaller polarizabilities and are sensitive to $\alpha$ variations. While the radioactive $^{227}$Th nuclear clock, with $Q=2\times10^{20}$, is still to be implemented for a direct laser excitation~\cite{NuclearTh-clock2022}, a HCI clock based on Ar$^{13+}$ with a modest $Q=4\times10^{13}$ has already been demonstrated via quantum logic spectroscopy~\cite{HCIclock-Ar13-2022}.  

One category of HCI candidates are based on hyperfine transitions of the $1s$ electron in hydrogenlike ions~\cite{HCIclock-H-2007,HCIclock-H-2014}. This transition has an energy of 6~$\mu$eV for the hydrogen atom, but scales up to 1.4~eV (886~nm) for ions as heavy as $^{207}$Pb$^{81+}$. The second variety of HCI clock transitions are from fine-structure splittings of $np$ valence electron systems~\cite{HCIclock-Al-2016,HCIclock-Sn-2016,HCIclock-Ar13-2020,HCIclock-Ca-2021}. Though these ions have polarizabilities many orders of magnitude smaller than neutral atoms or singly charged ions, their clock transitions are of magnetic dipole transitions with $Q\sim10^{13}$~\cite{HCIclock-H-2014,HCIclock-Ar13-2022}. Thus, one needs a long interrogation time to achieve high accuracy. Another type of HCI candidates comes from level crossings between the $4f-5s$~\cite{HCIclock-4f5s-2010,HCIclock-4f5s-2011,HCIclock-4f5s-2015},  $4f-5p$~\cite{HCIclock-4f5p-2014,HCIclock-4f5p-2019,HCIclock-4f5p-2021}, and $5f-6p$~\cite{HCIclock-5f6p-2012,HCIclock-5f6p-2015} orbitals. 
However, only a small number of ions close to the crossing points would provide optical clock transitions, not to mention that the close lying orbitals usually induce large polarizabilities~\cite{HCIclock-Pm-2016,HCIclock-Al-2016}.

\begin{figure}[b]
\includegraphics[width=0.45\textwidth]{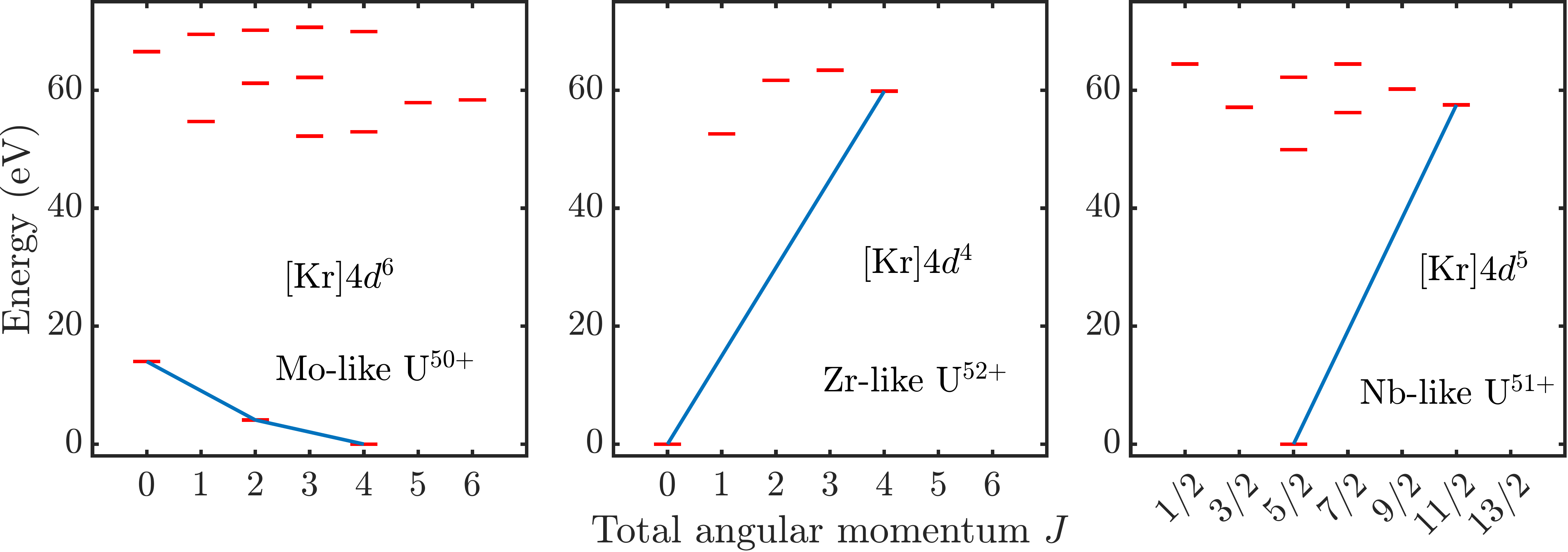}
\caption{\label{92U} Low-lying levels of (a) U$^{50+}$, (b) U$^{52+}$ and (c) U$^{51+}$ ions. Clock transitions are depicted by blue lines. Similar level structures exist for [Ar]$3d^m$ and [Xe]$4f^{14}5d^m$ ($m=4,5,6$) ions. 
} 
\end{figure}

In this Letter, we put forward a category of ultrastable HCI clocks that possess $Q$ factors beyond $10^{23}$ and electric polarizabilities in the range of $10^{-42}$~J~m$^2$/V$^2$. As shown in Fig.~\ref{92U}, the clock transitions come from fine-structure splittings in ions with one of the open-shell $nd^4$, $nd^5$ and $nd^6$ ground-state configurations. 
These configurations have tens of terms with total angular momenta ranging from $J=0$ and 1/2 to $J=6$ and 13/2.   
For light ions, the energy levels order linearly according to the values of the total angular momentum, and thus the lifetimes of the excited states are dominated by $M1$ transitions to $J\pm1$ states. These forbidden transitions have found numerious applications in the diagnostics of astrophysical and laboratory plasma properties~\cite{HCI-3d4-1991,HCI-3d4-1995,lopez2008visible}. However, for heavy ions which are also highly charged, the strong relativistic influence leads to a reordering of the terms and renders it possible to obtain ultrastable clocks based on high-multipolarity $E2/M3$ forbidden transitions.  

For the case of $nd^6$, optical clock transitions exist throughout the whole isoelectronic sequences up to U$^{68+}$, U$^{50+}$ and U$^{18+}$ for $n=3,4,5$, respectively, providing nearly a \textit{hundred} of HCI clocks. Some HCIs, such as U$^{68+}$, might be challenging to produce, but in principle can be obtained within an electron beam ion trap~\cite{HC-EBIT-2018} or a  HITRAP~\cite{hitrap-2015}.
High-precision spectroscopy of these transitions would enable an accurate determination of nuclear parameters for more than 70 elements and their isotopes ranging from \ce{_{22}^{46}Ti} to \ce{_{92}^{238}U} and beyond, and test state-of-the-art nuclear theories~\cite{NucPhys2022laser,NucPhys2022-Ni,NucPhys2022-Ca,NucPhys2020-Sn,Yordanov2020-Sn}.
For $nd^4$ and $nd^5$, tens of optical clock transitions also exist in light elements, but the transition energies scale up to the XUV and soft-x-ray region for heavy elements.  
Considering that the stability of optical lasers is approaching the fundamental limit set by thermal fluctuations of reference cavities~\cite{Localocsillator2004,Localocsillator2011,Localocsillator2017,Localocsillator2022}, clocks based on lasers with shorter wavelengths~\cite{XUVcomb-2013,XUVcomb-2014} would be one of the approaches to further improve the clock performance.
Furthermore, most of these ions also have more than two clock transitions with distinct sensitivities to the variation of $\alpha$, rendering them good candidates for the test of new physics beyond the standard model~\cite{newphysics-RMP2018}. 

\begin{figure}[t]
\includegraphics[width=0.45\textwidth]{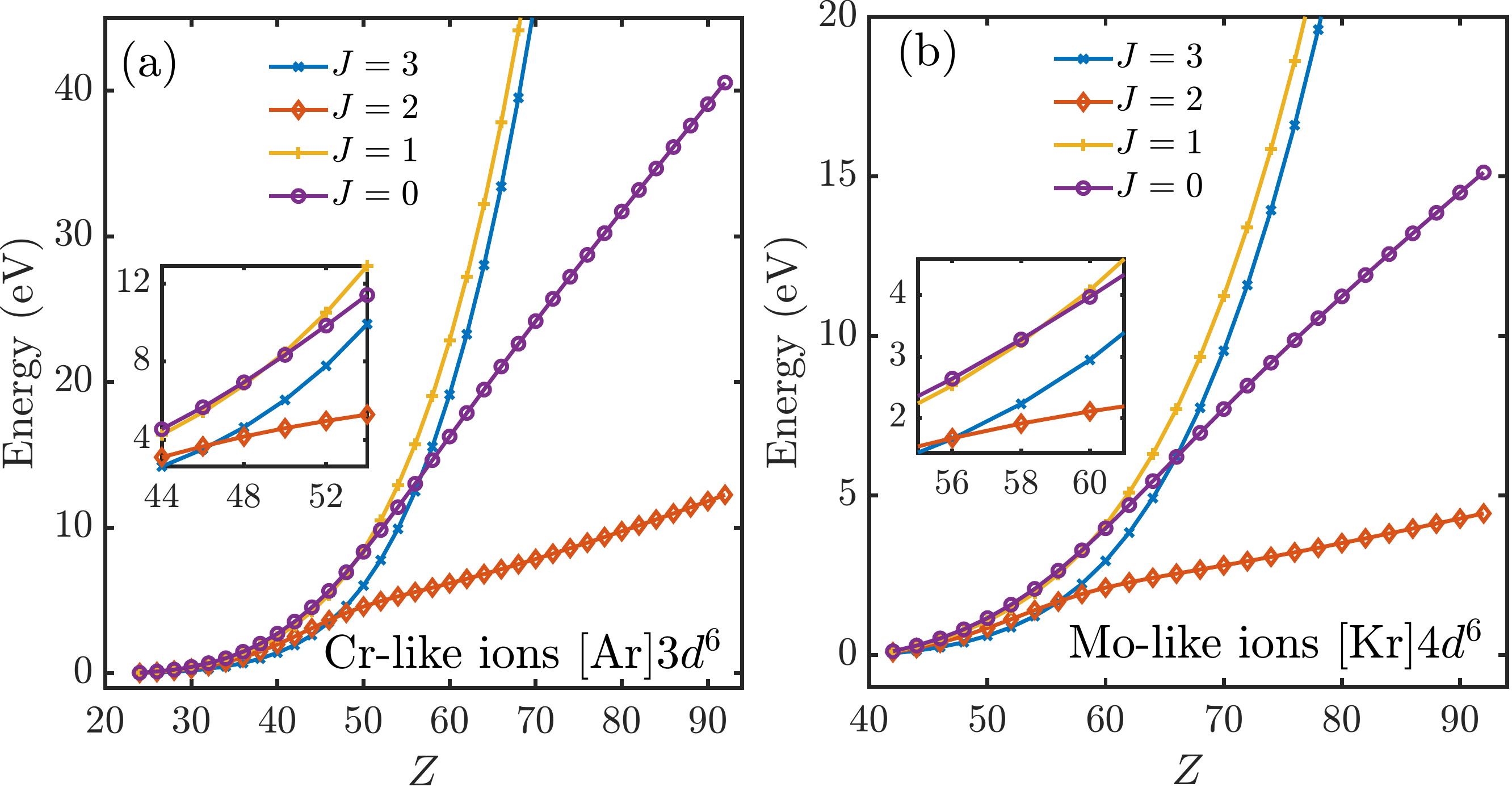}
\caption{\label{CrMo} Excitation energies of the $nd^6~{}^5D_{3,2,1,0}$ state in (a) Cr- and (b) Mo-like ions as functions of atomic number $Z$. With a ground state of $^5D_4$, term reordering leads to two clock states with $J=2$ and $J=0$, respectively. Insets are the enlarged depiction around the crossing points.} 
\end{figure}

\textit{MCDHF-RCI calculations} --- To accurately predict the transition energies and lifetimes of these clock states, we employ \textit{ab initio} fully relativistic multiconfiguration Dirac--Hartree--Fock (MCDHF) and configuration interaction (RCI) methods implemented in the GRASP2018 codes~\cite{grant2007relativistic,fischer2019grasp2018}. Within this scheme, the many-electron atomic state functions (ASFs) are expanded as a linear combination of configuration state functions (CSFs) which are $jj$-coupled Slater determinants of one-electron orbitals. The radial wavefunction for each orbital is obtained via solving the 
self-consistent MCDHF equations under the Dirac--Coulomb Hamiltonian. Then, the RCI method is applied to account for corrections arising from mass shift, quantum electrodynamic and Breit interactions.

 \begin{table}[t]
\renewcommand{\tablename}{Table}
\renewcommand{\thetable}{\arabic{table}}
\caption{Predicted transition energy $\hbar\omega$, lifetime $\tau$, sensitivity to $\alpha$ variations $K$, and differential polarizability $\Delta\alpha_{eg}$ ($10^{-40}$~J~m$^2$/V$^2$) of clock states in selected [Kr]$4d^6$ ions. }
\begin{ruledtabular}
\begin{tabular}{llllll}
ions   & $|e\rangle$ & $\hbar\omega$ (eV) & $\tau$ (s)  & $K$  & $\Delta\alpha_{eg}$\\
\hline
\multirow{2}{*}{Ce$^{16+}$}    & $^3D_2$ & 1.70(1) & 1156  & $-0.55$ & $-0.22$\\
   & $^3P_0$ & 3.23(1) & 602  & $-1.50$ & $-0.45$\\
\hline
\multirow{2}{*}{Dy$^{24+}$}    & $^3D_2$ & 2.20(1) & 545  & 0.06 & $-0.10$\\
   & $^3P_0$ & 5.98(2) & 11.7  & $-1.02$ & $-0.20$ \\
\hline
\multirow{2}{*}{W$^{32+}$}    & $^3D_2$ & 2.67(1) & 332  & 0.01 & $-0.05$ \\
   & $^3P_0$ & 8.69(3) & 1.63  & $-0.61$ & $-0.10$ \\
\hline
\multirow{2}{*}{Pb$^{40+}$}    & $^3D_2$ & 3.23(1) & 208  & 0.05 & $-0.031$ \\
   & $^3P_0$ & 11.25(4) & 0.58  & $-0.38$ & $-0.062$ \\
\hline
\multirow{2}{*}{U$^{50+}$}    & $^3D_2$ & 3.99(1) & 125  & 0.07 & $-0.017$\\
   & $^3P_0$ & 14.31(4) & 0.27  & $-0.23$ & $-0.035$ \\

\end{tabular}
\label{table:nd6}
\end{ruledtabular}
\end{table} 

\begin{figure}[t]
\includegraphics[width=0.45\textwidth]{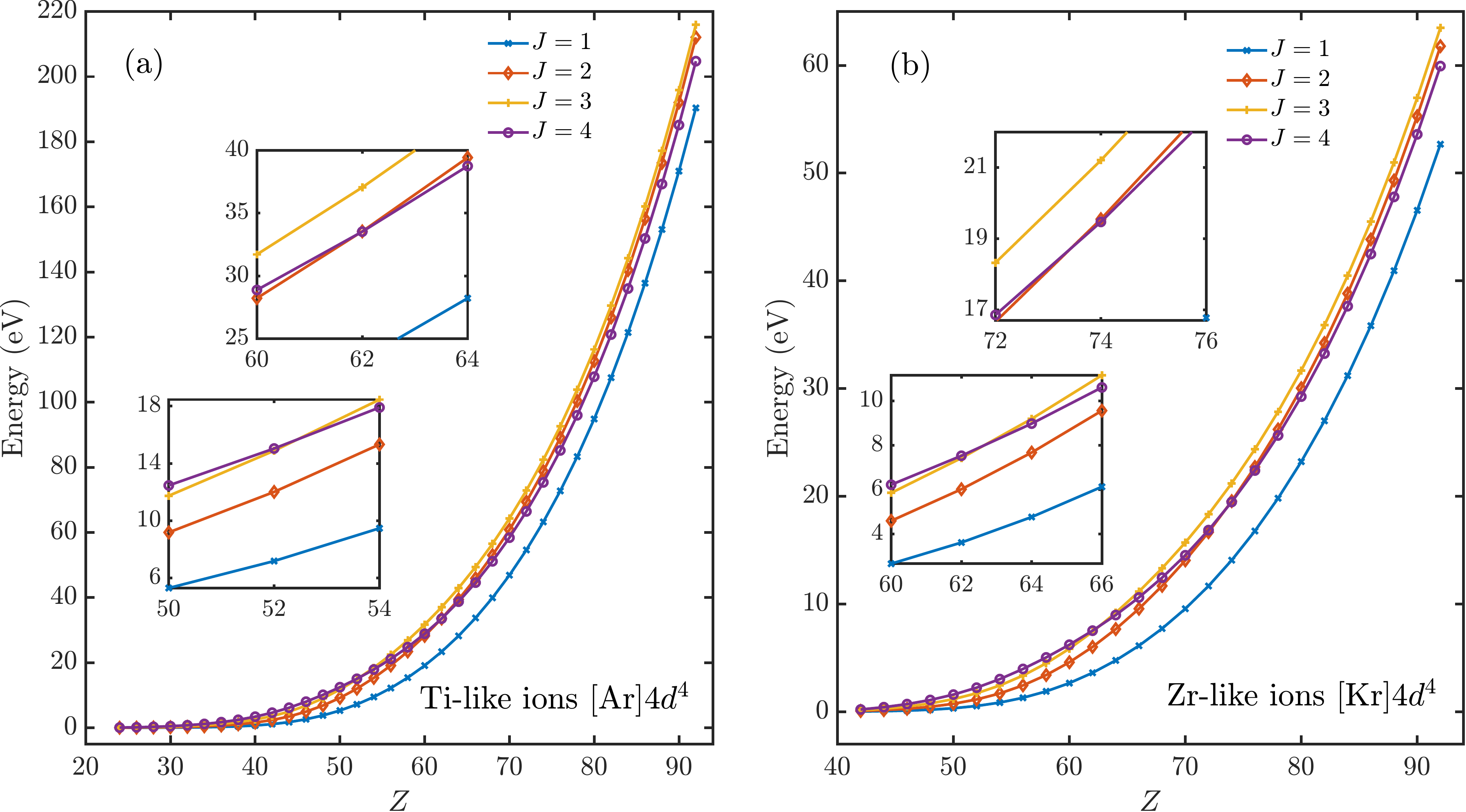}
\caption{\label{TiZr} Excitation energies of the $nd^4~{}^5D_{1,2,3,4}$ states in (a) Ti- and (b) Zr-like ions as functions of atomic number $Z$. Term reordering leads to long-lived $J=4$ state. Insets are the enlarged depiction around the crossing points.} 
\end{figure}

\begin{figure}[t]
\includegraphics[width=0.45\textwidth]{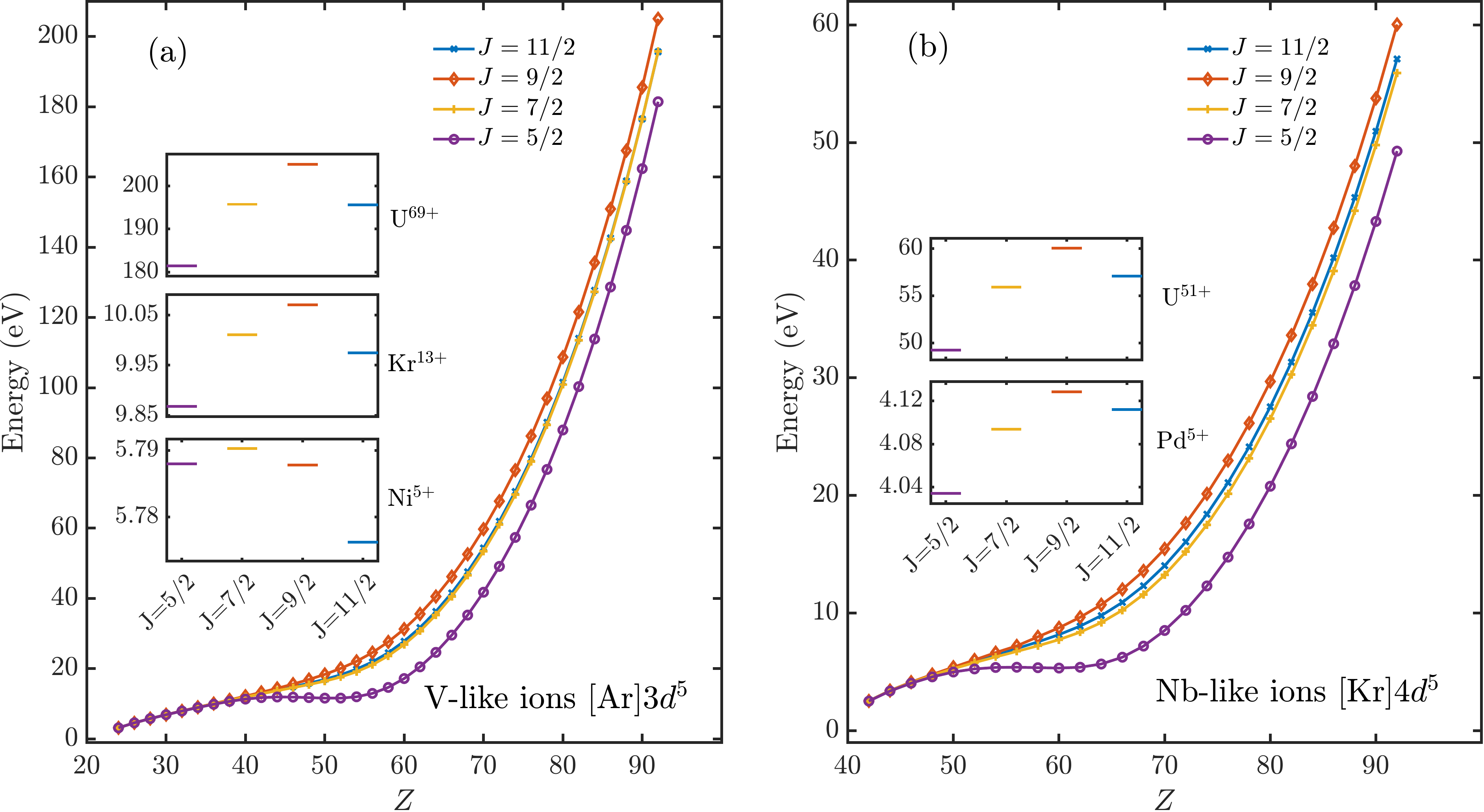}
\caption{\label{VNb} Excitation energies of the $nd^5~{}^4G_{5/2,7/2,9/2,11/2}$ states in (a) V- and (b) Nb-like ions as functions of atomic number $Z$. Term reordering leads to long-lived $J=11/2$ state. Insets are level structures of given ions.} 
\end{figure}

For the low-lying states in $nd^6$, $nd^4$ and $nd^5$, the scaling of their excitation energies with respect to the atomic number $Z$ are plotted in Figs.~\ref{CrMo}-\ref{VNb}, respectively, where a clear reordering of the levels is observed. These values are obtained via simple calculations that only account for correlation effects within the corresponding ground-state configurations. However, for specific ions where accurate values are needed, such as the ions around the crossing points discussed below and the ions shown in Table~\ref{table:nd6}, large-scale calculations with millions of CSFs are performed. This is obtained by allowing single and double excitations from the $ns$, $np$, and $nd$ orbitals up to the $8k$ orbitals. As a consequence, the calculated energies and lifetimes bear a relative accuracy of 0.5\% and 20\%, respectively~\cite{Kathrin2023}. Increasing the size of the CSF basis set systematically shifts the values of the excitation energies depicted in Figs.~\ref{CrMo}-\ref{VNb} downward by up to $10\%$, but has little effects on the crossing points of the levels.

\textit{$nd^6$ ions} --- For ions in this configuration, there are 34 atomic states with total angular momentum ranging from $J=0$ to 6, with a $^5D_4$ term being the ground state. In light ions, the corresponding low-lying excited states are $^5D_{3,2,1,0}$, respectively, and decay via $M1$ transitions. 
Usually, the energies of the fine-structure multiplet scale with $Z^2$~\cite{kozlov2018highly}, which is the case for $J=1,3$ shown in Fig.~\ref{CrMo}. However, for heavy elements, strong relativistic effects leads to large splitting between the $nd_{3/2}$ and $nd_{5/2}$ relativistic orbitals (the subscripts are the corresponding single-electron total angular momenta). As a consequence, the ASFs for states $J=0,2$ become dominated by the $nd_{3/2}^4nd_{5/2}^2$ configuration~\cite{suppl}, and their energies only increase linearly with respect to $Z$, and become the first two excited states. These two levels decay via slow $E2$ transitions and thus represent the two clock states illustrated in Fig.~\ref{92U}(a). We notice that a similar unusual scaling of the fine-structure splittings also allows clock transitions in low charged, so-called group-16-like ions~\cite{Ba4+-clock2020,Group16-clock2022,Group16-Ni12clock2023}.

For Mo-like ions ([Kr]$4d^6$) shown in Fig.~\ref{CrMo}(b), the level reordering of the $J=2$ state happens at Ba$^{14+}$. The transition energy between $^5D_4$ and $^3D_2$ is around 1.70~eV for Ce$^{16+}$, and scales up to 3.99~eV for U$^{50+}$. With a lifetime of 625 and 63~s for the two ions (see Table~\ref{table:nd6}), respectively, these transitions have a $Q$ factor of $1.6\times10^{18}$ and $3.8\times10^{17}$, respectively, which are orders of magnitude higher than those in current optical clocks~\cite{Al+-clock2019,Sr-lattice-clock2005,Cd-lattice-clock2019,In+-clock2019,Yb-lattice-clock2009,Hg-lattice-clock2012,Ca+-clock2016,Sr+-clock2004,Hg+-clock2001,CuAgAu-clock2021,Ra+-clock2022}. Particularly, they are more than 4 orders of magnitude higher than the first HCI clock based on Ar$^{13+}$~\cite{HCIclock-H-2014,HCIclock-Ar13-2020}. Thus, this isoelectronic system would contain more than 36 ion candidates suitable for optical HCI clocks operating at various wavelengths.

For Cr-like ions ([Ar]$3d^6$) as presented in Fig.~\ref{CrMo}(a), the $J=2$ metastable state starts to emerge from Pd$^{22+}$. As the electrons are more tightly bound, the transition energies are relatively larger than those in Mo-like ions but are still accessible for optical lasers. Specifically, the corresponding transition energy and lifetime are 3.78~eV and 437~s, respectively, for Cd$^{24+}$, and scale to 11.4~eV and 18.3~s, respectively, for U$^{68+}$. This corresponds to $Q$ factors around $10^{18}$ for more than 46 optical HCI clocks in this isoelectronic sequence.

Furthermore, the second metastable state (labelled by $J=0$) lies 3.23-14.3~eV and 6.75-38.6~eV above the ground state for Mo- and Cr-like ions, respectively. While it decays to the ground state via a highly forbidden $E4$ transition, its lifetime is dominated by an $E2$ transition to the $J=2$ excited state and ranges from 64~ms to 800~s. As shown in Table~\ref{table:nd6}, this $J=0$ state is much more sensitive to the variation of the fine-structure constant than that of the $J=2$ state. Taking the Mo-like Dy$^{24+}$ as an example, with the sensitivity defined by $K=(\Delta\omega/\omega)(\alpha/\Delta\alpha)$, one has a $K=0.07$ for the $^5D_4-{}^3D_2$ transition, but a $K=-1.02$ for the $^5D_4-{}^3P_0$ transition. Though the absolute sensitivities are smaller in comparison to some clock transitions near level crossings~\cite{HCIclock-4f5s-2010,HCIclock-4f5s-2011,HCIclock-4f5s-2015,HCIclock-4f5p-2014,HCIclock-4f5p-2019,HCIclock-4f5p-2021,HCIclock-5f6p-2012,HCIclock-5f6p-2015}, the distinct $\alpha$ disproportionality of two clock transitions in a single ion would largely reduce systematic uncertainties in detecting $\alpha$ variations. 
Similar dual-clock transitions also exist in Yb$^{+}$~\cite{alpha2004b,alpha2014}, which provides the current most stringent constraint of $\alpha$ variations. Nevertheless, one of the clock states in Yb$^{+}$ has a lifetime of only 7.3~ms ($Q=2.2\times10^{14}$) that limits further improvements.

Another advantage of these clock ions is that they are immune to external fields. This property is mainly characterized by the small static dipole polarizability defined by $\alpha_a=\sum_{k\neq a}-3 \hbar^2 f_{ka} /2m(\hbar\omega_{ka})^2$~\cite{starkShift2011}
Here, $|a\rangle$ is either the ground state $|g\rangle$  or the excited clock state $|e\rangle$ , $\hbar$ the reduced Planck constant, $m$ the electron rest mass, $f_{ka}$ the electric dipole oscillator strength of transition from $|a\rangle$ to an intermediate state $|k\rangle$, and $\hbar\omega_{ja}$ the corresponding transition energy in units of eV. The typical polarizability values for optical clocks are larger than $10^{-40}$~J~m$^2$/V$^2$~\cite{Polar2011-Ar+,Polar2016-Yb}. However, for ions like U$^{50+}$, the lowest $|k\rangle$ state is from the $4d^54f$ configuration which lies 270~eV above the two clock states. Making use of the Thomas--Reiche--Kuhn sum rule $\sum_{k\neq a}f_{ka}=N$ ($N$ being the number of electrons in the ion)~\cite{johnson2007atomic}, one can derive $\alpha_{e,g}<10^{-41}$~J~m$^2$/V$^2$ which are orders of magnitudes smaller in comparison to those in neutral atoms and singly charged ions. Accordingly, the differential polarizabilities $\Delta\alpha_{eg}=\alpha_{e}-\alpha_{g}$, as shown in Table~\ref{table:nd6}, are also much smaller.

\textit{$nd^4$ ions} --- As a counterpart of the above mentioned $nd^6$ isoelectronic systems, the $nd^4$ ions have similar spectroscopic terms but with a reversed level ordering. As shown in Fig.~\ref{TiZr}, these ions start with $^5D_0$ being the ground state and $^5D_{1,2,3,4}$ being the subsequent excited states for light elements. Though the energies of all the states scale dominantly with $Z^2$, the $^5D_4$ state gradually becomes lower than the $^5D_3$ and $^5D_2$ states and becomes a long-lived clock state shown in Fig.~\ref{92U}(b). Since their energies span from 9.0 to 200~eV, they can be probed with XUV frequency combs~\cite{XUVcomb-2013,XUVcomb-2014} and are suitable for future XUV and soft-x-ray ionic clocks. The higher transition energy would further increase the $Q$ factor to the range of $10^{22}$ and beyond, thus improve the short-term stability of a clock.

In the case of Zr-like ions ([Kr]$4d^4$), the clock transition energy varies from 9.0 eV in Gd$^{24+}$ to 60~eV in U$^{52+}$. Through Gd$^{24+}$ to Ta$^{33+}$, the $^3H_4$ state is still above the $^5D_2$ state. Therefore, its lifetime is mainly determined by the $E2$ transition and are of 5~hours to 20~days. With an energy of 9 to 20~eV, these clock transitions have a quality factor of $Q=10^{20}\sim10^{22}$. For heavier ions starting from W$^{34+}$, the $^3H_4$ state becomes lower than the $^5D_2$ state. Thus, the lifetime of this metastable state can be as long as tens to hundreds of years and the quality factor of the transitions would go far beyond $10^{22}$. Such a long lifetime makes excitation by lasers challenging, but the position of these states can be determined with a sub-eV accuracy through mass spectroscopy~\cite{HCImass-2020,Kathrin2023}. Furthermore, the application of twisted lasers could also enhance the transition rates of such high-multipolarity transitions~\cite{twisted2022}.

Similarly, for Ti-like ions with a configuration of [Ar]$3d^4$, the $^3H_4$ state lies between the $^5D_2$ and $^5D_3$ states for ions throughout Xe$^{32+}$ to Pm$^{39+}$, with a quality factor of $Q=10^{20}\sim10^{23}$ for transition energies between 17.9 and 33.0~eV. When the $^3H_4$ state becomes lower than the $^5D_2$ state in Sm$^{40+}$, its lifetime again becomes as long as hundreds of years. However, for the 202-eV metastable state in $_{92}$U$^{70+}$, its lifetime reduces to 48~days via an $E4$ spontaneous emission to the ground state.

\textit{$nd^5$ ions} --- For V- and Nb-like systems, the lowest few states in neutral V and Nb atoms are formed by the $nd^4(n+1)s$ and $nd^3(n+1)s^2$ configurations and have $M1$ forbidden transitions. However, for the singly charged Cr$^+$ and Mo$^+$ ions, the $nd^5~{}^6S_{5/2}$ term becomes the ground state. Nevertheless, the first excited state is still a long-lived $^6D_{1/2}$ state from the $nd^4(n+1)s$ configuration and represents a narrow near-infrared clock transition around 1.5~eV. Starting from V-like Mn$^{2+}$ and Nb-like Tc$^{2+}$, the low-lying excited states become $^4G_{11/2-5/2}$ from the $nd^5$ configuration. 

As shown in Fig.~\ref{VNb}, for light elements, the separations between these excited states are smaller than 10~meV that even the $M1$ decay channel are extremely slow. As a consequence, all of these states are ultrastable clock states. 
Especially, through V-like Mn$^{2+}$ to Br$^{12+}$ and Nb-like Tc$^{2+}$ to Ag$^{6+}$, our MCDHF-RCI calculations show that all the $nd^5~{}^4G$ terms have lifetimes longer than 1~s, with energies ranging from 3.0 to 9.4~eV. From Kr$^{13+}$ to Y$^{16+}$ and Cd$^{7+}$ to Sb$^{10+}$, even though the lifetime of the $^4G_{5/2}$ state becomes shorter than 1~s, the other three states still provide highly forbidden clock transitions for each ion in the range of 4.7 to 11.5~eV. These multiple optical clock transitions can be employed to infer the $\alpha$ variations and may also find applications in searching for a fifth force with a multidimensional King-plot~\cite{mKingplot2020}.

For elements heavier than V-like Mo$^{19+}$ and Nb-like Xe$^{13+}$, the energies of the excited states scale quadratically with the atomic number $Z$ (except for slightly abnormal behaviour of the $J=5/2$ state). The splittings between these states are far beyond 0.1~eV, thus fast $M1$ decay emerges. As a result, only the state with $J={11/2}$ that decays via a slow $E2/M3$ transition remains metastable. As shown in Fig.~\ref{92U}(c) for Nb-like U$^{51+}$, this clock state lies 57.2~eV above the ground state. With a lifetime of 8.3 days, it represents an XUV clock with $Q=6.3\times10^{22}$. In the case of V-like U$^{69+}$, the transition energy scales to 197~eV. With a lifetime of 12~hours, this soft-x-ray clock transition bears a quality factor of $Q=1.3\times10^{22}$.  

We note that, such a metastable state in Nb-like Pb$^{41+}$~\cite{Kathrin2023} has been detected in a recent high-precision mass spectroscopy experiment. The measured energy agrees with our predicted 31.68(13)-eV value obtained via large-scale MCDHF-RCI calculations. 
With a calculated lifetime of 26~days, it represents an XUV clock transition with $Q=1.1\times10^{23}$, and can be effectively probed with state-of-the-art XUV frequency combs~\cite{EUVclock-lyu,XUVcomb-2013,XUVcomb-2014,pupeza2021extreme,nauta2021xuv,zhang2020extreme,RamseyXUV2019}.


\textit{$5d^{4,5,6}$ ions} --- The above discussions also apply to \mbox{Hf-,} Ta-, and W-like ions through Hg$^{6+}$ to U$^{20+}$. Though most of them are radioactive, the energy and lifetime of their clock state varies from 1.06 to 11.0~eV and from 2~s to 500 days, respectively, with $Q=10^{16}-10^{23}$. Particularly, the dual clock states in W-like ions, with energies of 1.06 and 1.92~eV in Hg$^{6+}$, and 1.70 and 5.11~eV in U$^{18+}$, respectively, are also disproportional to $\alpha$ variations, with the sensitivity of the $J=0$ states being around $-0.61$ to $-1.23$. 

\textit{Conclusion} --- We have shown that there exist more than 100 ultrastable optical HCI clock candidates. They arise from fine-structure splittings in the open-shell $nd^4$, $nd^5$ and $nd^6$ ($n=3,4,5$) ions, for more than 70 elements and their isotopes, ranging from \ce{_{22}^{46}Ti} to \ce{_{92}^{238}U} and beyond. The corresponding transitions have $Q$ factors and polarizabilities many orders of magnitude larger and smaller, respectively, than for current optical clocks, and thus would allow for more accurate time keeping. Most of the ions have multiple clock states with opposite dependence on fine-structure constant variations. High-precision spectroscopy of these clock transitions would enrich the search for new physics beyond the standard model, as well as the test of state-of-the-art nuclear theories. Moreover, for heavy HCIs, these clock transitions also scale up to the XUV and soft-x-ray range, and could enable the construction of short-wavelength clocks. Similar clock transitions are also found in open-shell but more complicated $4f^{m}$ ions~\cite{4fions-lyu}.

\clearpage
\onecolumngrid
\section{Supplemental Materials}

This material summizes the mixing coefficients for low-lying states in selected $4d^6$, $4d^5$, and $4d^4$ ions. For heavy elements, the strong relativistic effect leads to large splittings between the $4d_{3/2}$ and $4d_{5/2}$ relativistic orbitals, with the subscripts being the corresponding single-electron total angular momenta. As the atomic number $Z$ increases, the energies of atomic states dominated by the $4d_{3/2}$ orbital scale much slower than those of the states involving high fraction of the $4d_{5/2}$ orbital. This impropotional scaling effect leads to the reordering of fine-structure terms, thus the emergence of metastable states. The notation $4d_{3/2}^2(0)$ means there are 2 electrons in the $4d_{3/2}$ orbital, and the total angular momentum of these two electrons is 0.

\subsection{$4d^6$ ions}\label{4d6}
For Mo and Mo-like Sn$^{8+}$, the ASFs are dominated by the $4d^3_{3/2}(3/2)4d^3_{5/2}(3/2)$ configuration where $d_{3/2}$ and $d_{5/2}$ orbitals are equally occupied. However, for U$^{50+}$, the energy of the $4d_{3/2}$ orbital is 55.8~eV lower than that of the $4d_{5/2}$ orbital. Thus, the dominate CSFs, for the low-lying $J=0,2,4$ states, are given by $4d^4_{3/2}(0)4d^2_{5/2}(0)$, $4d^4_{3/2}(0)4d^2_{5/2}(4)$, and $4d^4_{3/2}(0)4d^2(4)_{5/2}$, respectively. However, the $4d^4_{3/2}4d^2_{5/2}$ confituration does not allow angular coupling with $J=1,3$. Therefore, the terms with $J=1,3$ have more electrons occupying the $4d_{5/2}$ orbital, and thus will have much higher energy than the other three terms. Similar analysis also applies to the $4d^5$ and $4d^4$ ions shown in Sec.~\ref{4d5} and~\ref{4d4}, respectively. 

\begin{table}[h!]
\renewcommand{\tablename}{Table}
\renewcommand{\thetable}{\arabic{table}}
\caption{Absolute binding energies of the $4d_{3/2}$ and $4d_{5/2}$ relativistic orbitals in $4d^6$ ions. }
\begin{tabular}{l|l|l|l}
  \hline
  ion &  $E_{4d_{3/2}}$ (a.u.) & $E_{4d_{5/2}}$  (a.u.) & $\Delta E$ (eV)\\
   \hline
    42Mo$^{0+}$  &   0.214  &  0.207  &  0.190 \\
  50Sn$^{8+}$   &   5.916  & 5.866  &  1.361 \\
  92U$^{50+}$  &   107.599 &  105.548  &  55.811 \\
\label{tab.J0}
\end{tabular}
\end{table}

\begin{table}[h!]
\renewcommand{\tablename}{Table}
\renewcommand{\thetable}{\arabic{table}}
\caption{CSF components of the low-lying $J=0$ state. }
\begin{tabular}{l|l|l|l|l|l|l}
  \hline
  ion & ASF & \multicolumn{5}{c}{$c_k$} \\
   \hline
    42Mo$^{0+}$  &   0 +    &  0.76080 &   0.37843 &  -0.35227  &  0.28425  &  0.27035 \\
       &        &        3   &       1    &      2    &      5    &      4 \\
  50Sn$^{8+}$   &  0 +  &    0.74400 &   0.37938  &  0.36036  & -0.32716  &  0.25620 \\
      &        &         3   &       5   &       4  &        2   &       1 \\
  92U$^{50+}$  &   0 +   &   0.98227  &  0.16656 &  -0.07640 &  0.03946  &  0.00383 \\
       &       &         5   &       4   &       2   &       3   &       1 \\

      \hline
      \hline
  CSF &       &         1  &        2  &        3  &        4  &        5   \\
      &       &     $4d^6(0)_{5/2}$  &   $4d_{3/2}^2(0)4d^4_{5/2}(0)$    &    $4d_{3/2}^2(2)4d^4_{5/2}(2)$  &    $4d_{3/2}^3(3/2)4d^3_{5/2}(3/2)$  &  $4d_{3/2}^4(0)4d^2_{5/2}(0)$ 
\label{tab.J0}
\end{tabular}
\end{table}

\begin{table}[h!]
\renewcommand{\tablename}{Table}
\renewcommand{\thetable}{\arabic{table}}
\caption{CSF components of the state with $J=1$. }
\begin{tabular}{l|l|l|l|l|l|l}
  \hline
  ion & ASF  & \multicolumn{4}{c}{$c_k$} \\
   \hline
    42Mo$^{0+}$    &    1 +    &  0.64164 &  -0.51824 &  -0.41117 &  -0.38816 \\
          &       &      1     &     3     &     4     &     2 \\

  50Sn$^{8+}$   &   1 +   &   0.57585  &  0.50747 &  -0.50443  &  0.39550 \\
          &      &       3     &     4     &     1     &     2 \\
  92U$^{50+}$  &      1 +    &   0.79804   &  0.59318   &  0.09266   & -0.05178 \\
         &       &         4     &      3     &      2     &      1 \\

      \hline
      \hline
  CSF &       &         1  &        2  &        3  &        4 &  \\
      &       &        $4d_{3/2}(3/2)4d^5_{5/2}(5/2)$    &    $4d_{3/2}^2(2)4d^4_{5/2}(2)$  &    $4d_{3/2}^3(3/2)4d^3_{5/2}(5/2)$  &  $4d_{3/2}^3(3/2)4d^3_{5/2}(3/2)$ &
\label{tab.J1}
\end{tabular}
\end{table}

\begin{table}[h!]
\renewcommand{\tablename}{Table}
\renewcommand{\thetable}{\arabic{table}}
\caption{CSF components of the state with $J=2$. }
\begin{tabular}{l|l|l|l|l|l|l}
  \hline
  ion & ASF  & \multicolumn{5}{c}{$c_k$} \\
   \hline
    42Mo$^{0+}$    &    2 +   &   0.72324  &  0.37426  & -0.32648 &  -0.30319  & -0.27253\\
           &      &      5      &    1      &    8     &     6     &     4\\

  50Sn$^{8+}$   &   2 +   &   0.62622 &  -0.49708  & -0.35745 &   0.29182 &  -0.26834 \\
          &     &        5      &    8    &     6     &     1     &     4\\
  92U$^{50+}$  &      2 +   &   0.99339  &  0.09839  & -0.03938 &   0.03373 &  -0.02196\\
          &     &        8   &       6    &      2     &     7    &      5\\

      \hline
      \hline
  CSF &       &         1  &        2  &        3  &        4 &  \\
      &       &        $4d_{3/2}(3/2)4d^5_{5/2}(5/2)$    &    $4d_{3/2}^2(0)4d^4_{5/2}(2)$  &     $4d_{3/2}^2(2)4d^4_{5/2}(0)$  &   $4d_{3/2}^2(2)4d^4_{5/2}(2)$ & \\
   &       &         5  &       6  &        7  &        8 &  \\
      &       &         $4d_{3/2}^2(2)4d^4_{5/2}(4)$    &    $4d_{3/2}^3(3/2)4d^3_{5/2}(5/2)$  &    $4d_{3/2}^3(3/2)4d^3_{5/2}(3/2)$  &  $4d_{3/2}^4(0)4d^2_{5/2}(2)$ & \\
\label{tab.J2}
\end{tabular}
\end{table}

\begin{table}[h!]
\renewcommand{\tablename}{Table}
\renewcommand{\thetable}{\arabic{table}}
\caption{CSF components of the state with $J=3$. }
\begin{tabular}{l|l|l|l|l|l|l}
  \hline
  ion & ASF  & \multicolumn{5}{c}{$c_k$} \\
   \hline
    42Mo$^{0+}$    &    3 +  &    0.74515 &  -0.56412  &  0.22897 &  -0.20396 &  -0.14349\\
          &         &    6    &      3   &       4    &      1     &     5\\

  50Sn$^{8+}$   &   3 +    &  0.74515 &  -0.56412  &  0.22897 &  -0.20396  & -0.14349\\
          &     &        6     &     3      &    4     &     1    &      5\\
  92U$^{50+}$  &      3 +   &   0.89099  &  0.32023  & -0.30164  & -0.10551  &  0.02863 \\
          &        &     6    &      4    &      5     &     3    &      2\\

      \hline
      \hline
  CSF &       &         1  &        2  &        3  &        4 &  \\
      &       &        $4d_{3/2}(3/2)4d^5_{5/2}(5/2)$    &    $4d_{3/2}^2(2)4d^4_{5/2}(2)$  &     $4d_{3/2}^2(2)4d^4_{5/2}(4)$  &   $4d_{3/2}^3(3/2)4d^3_{5/2}(5/2)$ & \\
   &       &         5  &       6  &          &         &  \\
      &       &         $4d_{3/2}^3(3/2)4d^3_{5/2}(3/2)$    &    $4d_{3/2}^3(3/2)4d^3_{5/2}(9/2)$  &       &    & \\
\label{tab.J3}
\end{tabular}
\end{table}

\begin{table}[h!]
\renewcommand{\tablename}{Table}
\renewcommand{\thetable}{\arabic{table}}
\caption{CSF components of the state with $J=4$. }
\begin{tabular}{l|l|l|l|l|l|l}
  \hline
  ion & ASF  & \multicolumn{5}{c}{$c_k$} \\
   \hline
    42Mo$^{0+}$    &    4 +   &   0.67191 &   0.47875 &  -0.40263 &  -0.29717 &  -0.18941\\
            &      &     6    &      7    &      4    &      2      &    1\\

  50Sn$^{8+}$   &   4 +   &   0.65905 &   0.59695 &   -0.31364 &  -0.26429 &   0.13655\\
           &     &       6    &      7      &    4    &      2  &       5\\
  92U$^{50+}$  &      4 +   &   0.99493  &  0.08798  & -0.04455 &  -0.01577  & -0.01165\\
           &     &       7   &       6    &      2  &        4 &         3\\

      \hline
      \hline
  CSF &       &         1  &        2  &        3  &        4 &  \\
      &       &        $4d_{3/2}(3/2)4d^5(5/2)$    &    $4d_{3/2}^2(0)4d_{5/2}^4(4)$  &     $4d_{3/2}^2(2)4d_{5/2}^4(2)$  &   $4d_{3/2}^2(2)4d_{5/2}^4(4)$ & \\
   &       &         5  &       6  &      7    &         &  \\
      &       &         $4d_{3/2}^3(3/2)4d_{5/2}^3(5/2)$    &    $4d_{3/2}^3(3/2)4d_{5/2}^3(9/2)$  &    $4d_{3/2}^4(0)4d_{5/2}^2(4)$   &    & \\
\label{tab.J4}
\end{tabular}
\end{table}

~\clearpage
\subsection{$4d^5$ ions}\label{4d5}

\begin{table}[h!]
\renewcommand{\tablename}{Table}
\renewcommand{\thetable}{\arabic{table}}
\caption{Absolute binding energies of the $4d_{3/2}$ and $4d_{5/2}$ relativistic orbitals in $4d_{5/2}^5$ ions. }
\begin{tabular}{l|l|l|l}
  \hline
  ion &  $E_{4d_{3/2}}$ (a.u.) & $E_{4d_{5/2}}$  (a.u.) & $\Delta E$ (eV)\\
   \hline
    42Mo$^{0+}$  &   0.554  &  0.547  &  0.190 \\
  50Sn$^{8+}$   &   6.758  & 6.706  &  1.145 \\
  92U$^{50+}$  &   110.328 &  108.162  &  58.940 \\
\label{tab.J0}
\end{tabular}
\end{table}

\begin{table}[h!]
\renewcommand{\tablename}{Table}
\renewcommand{\thetable}{\arabic{table}}
\caption{CSF components of the ground state with $J=5/2$. }
\begin{tabular}{l|l|l|l|l|l|l}
  \hline
  ion & ASF  & \multicolumn{5}{c}{$c_k$} \\
   \hline
    42Mo$^{+}$  &   5/2 + &   0.61759  &  0.50663 &  -0.45259  &  0.21928 &  -0.19577\\
           &        &     7    &      9    &      3     &    10     &     4 \\
  50Sn$^{9+}$   &  5/2 +  &    0.58952  &  0.57196 &  -0.37560  &  0.29939 &  -0.19366 \\
           &       &      7   &       9    &      3    &     10   &       4 \\
  92U$^{51+}$  &   5/2 +   &   0.99305  &  0.09767 &  -0.05647 &   0.02543 &   0.01799 \\
         &         &     10    &      9       &   4     &     7   &       6 \\

      \hline
      \hline
  CSF &       &         1  &        2  &        3  &        4 & 5 \\
      &       &        $4d_{5/2}^5(5/2)$    &    $4d_{3/2}(3/2)4d_{5/2}^4(2)$  &    $4d_{3/2}(3/2)4d_{5/2}^4(4)$  &  $4d_{3/2}^2(0)4d_{5/2}^3(5/2)$ & $4d_{3/2}^2(2)4d_{5/2}^3(5/2)$ \\
   &       &         6  &        7  &        8 &      9 & 10 \\
      &       &        $4d_{3/2}^2(2)4d_{5/2}^3(3/2)$    &    $4d_{3/2}^2(2)4d_{5/2}^3(9/2)$  &    $4d_{3/2}^3(3/2)4d_{5/2}^2(2)$  &  $4d_{3/2}^3(3/2)4d_{5/2}^2(4)$ & $4d_{3/2}^4(0)4d_{5/2}(5/2)$ \\
\label{tab.J0}
\end{tabular}
\end{table}

\begin{table}[h!]
\renewcommand{\tablename}{Table}
\renewcommand{\thetable}{\arabic{table}}
\caption{CSF components of the excited state with $J=5/2$. }
\begin{tabular}{l|l|l|l|l|l|l}
  \hline
  ion & ASF  & \multicolumn{5}{c}{$c_k$} \\
   \hline
    42Mo$^{+}$    &    5/2 +   & 0.49676  & -0.43361 &   0.35679 &  -0.31041 &   0.29820\\
          &        &       6     &     8     &     2     &     9    &      7 \\

  50Sn$^{9+}$   &   5/2 +   &   0.68269  &  0.39504  & -0.36717  &  0.27866 &  -0.24829 \\
             &      &    10      &    6   &       8   &       3     &     9 \\
  92U$^{51+}$  &      5/2 +    &   0.87733  &  0.45249   & 0.11797  & -0.08731 &  -0.05064 \\
             &        &  9     &     8     &     7    &     10   &       3  \\

      \hline
      \hline
  CSF &       &         1  &        2  &        3  &        4 & 5 \\
      &       &        $4d_{5/2}^5(5/2)$    &    $4d_{3/2}(3/2)4d_{5/2}^4(2)$  &    $4d_{3/2}(3/2)4d_{5/2}^4(4)$  &  $4d_{3/2}^2(0)4d_{5/2}^3(5/2)$ & $4d_{3/2}^2(2)4d_{5/2}^3(5/2)$ \\
   &       &         6  &        7  &        8 &      9 & 10 \\
      &       &        $4d_{3/2}^2(2)4d_{5/2}^3(3/2)$    &    $4d_{3/2}^2(2)4d_{5/2}^3(9/2)$  &    $4d_{3/2}^3(3/2)4d_{5/2}^2(2)$  &  $4d_{3/2}^3(3/2)4d_{5/2}^2(4)$ & $4d_{3/2}^4(0)4d_{5/2}(5/2)$ \\
\label{tab.J1}
\end{tabular}
\end{table}

\begin{table}[h!]
\renewcommand{\tablename}{Table}
\renewcommand{\thetable}{\arabic{table}}
\caption{CSF components of the excited state with $J=7/2$. }
\begin{tabular}{l|l|l|l|l|l|l}
  \hline
  ion & ASF  & \multicolumn{5}{c}{$c_k$} \\
   \hline
    42Mo$^{+}$    &    7/2 +   &   0.59145 &  -0.47959 &   0.39208 &  -0.32789  &  0.28363\\
             &      &     5    &      7   &       2     &     6    &      4 \\

  50Sn$^{9+}$   &   7/2 +   &   0.65387 &  -0.46013  &  0.38918 &  -0.31272 &  -0.24884 \\
           &        &     7   &       5   &       6   &       2    &      4 \\
  92U$^{51+}$  &      7/2 +   &   0.92695 &   0.37150 &  -0.03240 &  -0.02816 &  -0.02348 \\
           &        &     7     &     6      &    2   &       4      &    1 \\

      \hline
      \hline
  CSF &       &         1  &        2  &        3  &        4 & 5 \\
      &       &        $4d_{3/2}(3/2)4d_{5/2}^4(2)$    &    $4d_{3/2}(3/2)4d_{5/2}^4(4)$  &     $4d_{3/2}^2(2)4d_{5/2}^3(5/2)$  &   $4d_{3/2}^2(2)4d_{5/2}^3(3/2)$ &          $4d_{3/2}^2(2)4d_{5/2}^3(9/2)$\\
   &       &                6  &        7  &         & & \\
      &           &   $4d_{3/2}^3(3/2)4d_{5/2}^2(2)$  &    $4d_{3/2}^3(3/2)4d_{5/2}^2(4)$   &  & & \\
\label{tab.J2}
\end{tabular}
\end{table}

\begin{table}[h!]
\renewcommand{\tablename}{Table}
\renewcommand{\thetable}{\arabic{table}}
\caption{CSF components of the state with $J=9/2$. }
\begin{tabular}{l|l|l|l|l|l|l}
  \hline
  ion & ASF  & \multicolumn{5}{c}{$c_k$} \\
   \hline
    42Mo$^{+}$    &    9/2 +  &    0.81369 &  -0.44351  &  0.37558 &   0.01020  &  0.00565 \\
           &        &    4    &      5    &      1    &      2     &     3 \\

  50Sn$^{9+}$   &   9/2 +    &  0.77477  & -0.55675  &  0.29631 &   0.03996 &   0.01902 \\
            &       &    4      &    5    &      1  &       2    &      3 \\
  92U$^{51+}$  &      9/2 +   &   0.99487 &  -0.08321 &  -0.04440 &  -0.02742 &  -0.02423 \\
         &         &     5      &    4      &    2   &       3  &       1 \\

      \hline
      \hline
  CSF &       &         1  &        2  &        3  &        4 & 5 \\
      &       &        $4d_{3/2}(3/2)4d_{5/2}^4(4)$    &    $4d_{3/2}^2(0)4d_{5/2}^3(9/2)$  &     $4d_{3/2}^2(2)4d_{5/2}^3(5/2)$  &   $4d_{3/2}^2(2)4d_{5/2}^3(9/2)$ & $4d_{3/2}^3(3/2)4d_{5/2}^2(4)$  \\
\label{tab.J3}
\end{tabular}
\end{table}

\begin{table}[h!]
\renewcommand{\tablename}{Table}
\renewcommand{\thetable}{\arabic{table}}
\caption{CSF components of the state with $J=11/2$. }
\begin{tabular}{l|l|l|l|l}
  \hline
  ion & ASF  & \multicolumn{3}{c}{$c_k$} \\
   \hline
    42Mo$^{+}$    &    11/2 +   &   0.71539 &  -0.53453  &  0.44999 \\
           &       &      2       &   3      &    1 \\

  50Sn$^{9+}$   &   11/2 +   &   0.67722  & -0.64170 &  -0.35999\\
          &       &      3    &      2     &     1 \\
  92U$^{51+}$  &      11/2 +   &   0.99809  & -0.04816 &  -0.03862 \\
            &        &   3    &      2       &   1\\

      \hline
      \hline
  CSF &       &         1  &        2  &        3      \\
      &       &        $4d_{3/2}(3/2)4d_{5/2}^4(4)$    &    $4d_{3/2}^2(2)4d_{5/2}^3(9/2)$  &     $4d_{3/2}^3(3/2)4d_{5/2}^2(4)$ 
\label{tab.J4}
\end{tabular}
\end{table}

~\clearpage
\subsection{$4d^4$ ions}\label{4d4}

\begin{table}[h!]
\renewcommand{\tablename}{Table}
\renewcommand{\thetable}{\arabic{table}}
\caption{Absolute binding energies of the $4d_{3/2}$ and $4d_{5/2}$ relativistic orbitals in $4d_{5/2}^4$ ions. }
\begin{tabular}{l|l|l|l}
  \hline
  ion &  $E_{4d_{3/2}}$ (a.u.) & $E_{4d_{5/2}}$  (a.u.) & $\Delta E$ (eV)\\
   \hline
    42Mo$^{0+}$  &   0.968  &  0.960  &  0.218 \\
  50Sn$^{8+}$   &   7.628  & 7.574  &  1.469 \\
  92U$^{50+}$  &   113.079 &  110.888  &  59.620 \\
\label{tab.J0}
\end{tabular}
\end{table}

\begin{table}[h!]
\renewcommand{\tablename}{Table}
\renewcommand{\thetable}{\arabic{table}}
\caption{CSF components of the state with $J=0$. }
\begin{tabular}{l|l|l|l|l|l|l}
  \hline
  ion & ASF  & \multicolumn{5}{c}{$c_k$} \\
   \hline
    42Mo$^{2+}$  &   0 + &   0.72165  &  0.51071 &  -0.36020   & 0.22064 &   0.19994\\
           &        &    4      &    5      &    3      &    1   &       2 \\
  50Sn$^{10+}$   &  0 +  &    0.71207   & 0.58598 &  -0.34109 &   0.14457 &   0.11108\\
           &       &     5     &     4       &   3      &    1      &    2 \\
  92U$^{52+}$  &   0 +   &   0.99720 &  -0.06621  &  0.03466  &  0.00303  & -0.00088\\
         &         &     5      &    3      &    4      &    1      &    2 \\

      \hline
      \hline
  CSF &       &         1  &        2  &        3  &        4  &        5   \\
      &       &     $4d_{5/2}^4(0)$  &   $4d_{3/2}(3/2)4d_{5/2}^3(3/2)$    &    $4d_{3/2}^2(0)4d_{5/2}^2(0)$  &    $4d_{3/2}^2(2)4d_{5/2}^2(2)$  &  $4d_{3/2}^4(0)$ 
\label{tab.J0}
\end{tabular}
\end{table}

\begin{table}[h!]
\renewcommand{\tablename}{Table}
\renewcommand{\thetable}{\arabic{table}}
\caption{CSF components of the state with $J=1$. }
\begin{tabular}{l|l|l|l|l|l|l}
  \hline
  ion & ASF  & \multicolumn{4}{c}{$c_k$} \\
   \hline
    42Mo$^{2+}$    &    1 +   & 0.74455  & -0.45667  &  0.35526 &   0.33301\\
          &        &     4   &       1    &      3      &    2 \\

  50Sn$^{10+}$   &   1 +   &   0.84828  & -0.36950  &  0.29390 &  0.23983\\
             &      &    4      &    1     &     3      &    2 \\
  92U$^{52+}$  &      1 +    &   0.99761 &  -0.05523  &  0.03652  &  0.01960\\
             &        &  4    &      1      &    3      &    2 \\

      \hline
      \hline
  CSF &       &         1  &        2  &        3  &        4 &  \\
      &       &        $4d_{3/2}(3/2)4d_{5/2}^3(5/2)$    &    $4d_{3/2}(3/2)4d_{5/2}^3(3/2)$  &    $4d_{3/2}^2(2)4d_{5/2}^2(2)$  &  $4d_{3/2}^3(3/2)4d_{5/2}(5/2)$ &
\label{tab.J1}
\end{tabular}
\end{table}

\begin{table}[h!]
\renewcommand{\tablename}{Table}
\renewcommand{\thetable}{\arabic{table}}
\caption{CSF components of the state with $J=2$. }
\begin{tabular}{l|l|l|l|l|l|l}
  \hline
  ion & ASF  & \multicolumn{5}{c}{$c_k$} \\
   \hline
    42Mo$^{2+}$    &    2 +   &   0.75176  & -0.42854 &   0.26385  & -0.26042 &   0.24193\\
             &      &    7      &    8      &    2    &      6    &      1\\

  50Sn$^{10+}$   &   2 +   &   0.75312  & -0.50378 &  -0.23758  &  0.22462  & -0.18064\\
           &        &    7     &     8     &     6      &    2       &   4\\
  92U$^{52+}$  &      2 +   &   0.98830 &  -0.14298 &  -0.04338 &   0.02404  &  0.01494\\
           &        &    8    &      7      &    2    &      5       &   3\\

      \hline
      \hline
  CSF &       &         1  &        2  &        3  &        4 &  \\
      &       &        $4d_{5/2}^4(2)$    &    $4d_{3/2}(3/2)4d_{5/2}^(5/2)$  &     $4d_{3/2}(3/2)4d_{5/2}^(3/2)$  &   $4d_{3/2}^2(0)4d_{5/2}^2(2)$ & \\
   &       &         5  &       6  &        7  &        8 &  \\
      &       &         $4d_{3/2}^2(2)4d_{5/2}^2(0)$    &   $4d_{3/2}^2(2)4d_{5/2}^2(2)$  &    $4d_{3/2}^2(2)4d_{5/2}^2(4)$  &  $4d_{3/2}^3(3/2)4d_{5/2}(5/2)$ & \\
\label{tab.J2}
\end{tabular}
\end{table}

\begin{table}[h!]
\renewcommand{\tablename}{Table}
\renewcommand{\thetable}{\arabic{table}}
\caption{CSF components of the state with $J=3$. }
\begin{tabular}{l|l|l|l|l|l|l}
  \hline
  ion & ASF  & \multicolumn{5}{c}{$c_k$} \\
   \hline
    42Mo$^{2+}$    &    3 +  &    0.67715 &  -0.62242 &  -0.31231 &   0.19152  & -0.10257\\
           &        &    5      &    3       &   6     &     1   &       4\\

  50Sn$^{10+}$   &   3 +    &  0.71626 &  -0.53100 &  -0.40474  &  0.17041 &  -0.08183\\
            &       &   5      &    3     &     6      &    1      &    4\\
  92U$^{52+}$  &      3 +   &   0.99140  & -0.11531 &  -0.04676 &  -0.03413  &  0.01854\\
         &         &     6    &      5      &    4     &     1      &    3\\

      \hline
      \hline
  CSF &       &         1  &        2  &        3  &        4 &  \\
      &       &        $4d_{3/2}(3/2)4d_{5/2}^3(5/2)$    &    $4d_{3/2}(3/2)4d_{5/2}^3(3/2)$  &     $4d_{3/2}(3/2)4d_{5/2}^3(9/2)$  &   $4d_{3/2}^2(2)4d_{5/2}^2(2)$ & \\
   &       &         5  &       6  &          &         &  \\
      &       &         $4d_{3/2}^2(2)4d_{5/2}^2(4)$    &    $4d_{3/2}^3(3/2)4d_{5/2}(5/2)$  &       &    & \\
\label{tab.J3}
\end{tabular}
\end{table}

\begin{table}[h!]
\renewcommand{\tablename}{Table}
\renewcommand{\thetable}{\arabic{table}}
\caption{CSF components of the state with $J=4$. }
\begin{tabular}{l|l|l|l|l|l|l}
  \hline
  ion & ASF  & \multicolumn{5}{c}{$c_k$} \\
   \hline
    42Mo$^{2+}$    &    4 +   &   0.64516 &  -0.46912 &  -0.39396 &   0.32122   & 0.25876\\
           &       &     3      &    6     &     1     &     4       &   7\\

  50Sn$^{10+}$   &   4 +   &   0.55959 &  -0.52906  &  0.40707  &  0.33947 &  -0.28611\\
          &       &      3      &    6     &     7    &      4     &     1\\
  92U$^{52+}$  &      4 +   &   0.99728 &  -0.04564  &  0.04427 &  -0.02746 &  0.01816\\
            &        &   7       &   2     &     4    &      6       &   5\\

      \hline
      \hline
  CSF &       &         1  &        2  &        3  &        4 &  \\
      &       &        $4d_{5/2}^4(4)$    &    $4d_{3/2}(3/2)4d_{5/2}^3(5/2)$  &     $4d_{3/2}(3/2)4d_{5/2}^3(9/2)$  &   $4d_{3/2}^2(2)4d_{5/2}^2(4)$ & \\
   &       &         5  &       6  &      7    &         &  \\
      &       &         $4d_{3/2}^2(2)4d_{5/2}^2(2)$    &   $4d_{3/2}^2(2)4d_{5/2}^2(4)$  &    $4d_{3/2}^3(3/2)4d_{5/2}(5/2)$   &    & \\
\label{tab.J4}
\end{tabular}
\end{table}

\end{document}